\iffalse
\documentclass[preprint]{aastex} 

\else
\documentclass[iop]{emulateapj}
\usepackage{apjfonts, epsfig}
\clearpage
\slugcomment{Accepted for publication in PASP}
\slugcomment{}

\fi

\shorttitle{CRBLASTER}
\shortauthors{Mighell}

\begin{document}

\title{CRBLASTER: A Parallel-Processing Computational Framework for\break
Embarrassingly-Parallel Image-Analysis Algorithms}

\author{Kenneth John Mighell}
\affil{National Optical Astronomy Observatory, 950 North Cherry Avenue, Tucson, AZ 85719\slugcomment{}}

\begin{abstract}
The development of parallel-processing image-analysis codes is generally a challenging task 
that requires complicated choreography of interprocessor communications.
If, however, the image-analysis algorithm is embarrassingly parallel,
then the development of a parallel-processing implementation of that algorithm
can be a much easier task to accomplish because, by definition, there is little need
for communication between the compute processes.
I describe the design, implementation, and performance of a
parallel-processing image-analysis application, called {\sc{crblaster}},
which does cosmic-ray rejection of CCD (charge-coupled device) images
using the embarrassingly-parallel \mbox{\sc{l.a.cosmic}} algorithm.
{\sc{crblaster}} is written in C using the  high-performance computing
industry standard Message Passing Interface (MPI) library. 
{\sc{crblaster}} uses a two-dimensional image partitioning algorithm
which partitions an input image into $N$ rectangular subimages of nearly equal area;
the subimages include sufficient additional pixels along common image partition edges
such that the need for communication between compute processes is eliminated.
The code has been designed to be used 
by research scientists who are familiar with C
as a parallel-processing computational framework
that enables the easy development of parallel-processing image-analysis programs
based on embarrassingly-parallel algorithms.
The {\sc{crblaster}} source code is freely available at the official application
website at the National Optical Astronomy Observatory.
Removing cosmic rays from a single 800$\times$800 pixel {\sl{Hubble Space Telescope}} 
WFPC2 image takes 44 seconds with the 
IRAF script {\small\tt{lacos\_im.cl}} running on a single core of 
an Apple Mac Pro computer with two 2.8-GHz quad-core Intel Xeons processors.
{\sc{crblaster}} is 7.4 times faster processing the same image on a single core on the same
machine.
Processing the same image with {\sc{crblaster}}
simultaneously on all 8 cores of the same machine 
takes 0.875 seconds --- which is a speedup factor of 50.3 times faster than the IRAF script.
A detailed analysis is presented 
of the performance of {\sc{crblaster}} using between 1 and 57 processors
on a low-power Tilera 
700-MHz 64-core TILE64 processor.
\end{abstract}

\slugcomment{Accepted for publication in PASP}

\section{INTRODUCTION}

An image-analysis algorithm can be classified as being embarrassingly parallel if it can be parallelized
by simply breaking up the image into many subimages that can processed individually without
requiring any communication between the computation processes \citep[e.g.,][]{WilkinsonAllen2004}\,.
Astrophysical image-analysis applications are
excellent candidates for embarrassingly-parallel computation 
if their analysis of any given subimage 
does not affect the analysis of another subimage.

In this article I will describe 
a parallel-processing image-analysis application, called {\sc{crblaster}},
which does cosmic-ray rejection of CCD (charge-coupled device) images
using the embarrassingly-parallel  \mbox{\sc{l.a.cosmic}} algorithm \citep{vanDokkum2001}.
The code has been designed to be used 
by research scientists who are familiar with C computer language 
\citep{KernighanRitchie1988}
as a parallel-processing computational framework
that enables the easy development of parallel-processing image-analysis programs
based on embarrassingly-parallel algorithms.
I describe the various sources of ``cosmic ray'' radiation in \S 2 and then
discuss the cosmic-ray degradation of CCD images in \S 3. I discuss several techniques
to remove cosmic rays in astronomical CCD observations in \S 4.
The embarrassingly-parallel \mbox{\sc{l.a.cosmic}} algorithm is described in \S 5.
The design, implementation, and performance of {\sc{crblaster}} on two different computing
platforms is presented in \S 6.
The article concludes with a discussion in \S 7.\break

\section{SOURCES OF ``COSMIC-RAY'' RADIATION}

In deep space, cosmic rays are energetic particles of extraterrestrial origin.
About 90\% of cosmic rays are protons and about 9\% are alpha particles (helium nuclei).
While most cosmic rays come from outside of the solar system, solar flares are a 
significant source of low-energy cosmic rays.

The {\em{Spitzer Space Telescope}} 
is on a trailing-Earth heliocentric orbit around the Sun
and is currently about 0.7 AU behind the earth.  
The {\em{Hubble Space Telescope}} {\em{(HST)}} circles the Earth every $\sim$96 minutes 
on a low-Earth orbit of $\sim$350 miles above the surface of the planet.
{\em{Spitzer}} encounters much more cosmic-ray radiation
than {\em{HST}} because it is far beyond the protection of the Earth's magnetic field;
the worst damage occurs during head-on encounters with coronal mass ejections 
(see, e.g., Fig.~7.19 of the 
{\em{IRAC Instrument Handbook}}\footnote{The {\em{IRAC Instrument Handbook}},
(Version 1.0) is currently available at\newline
http://ssc.spitzer.caltech.edu/irac/iracinstrumenthandbook/
}).

The South Atlantic Anomaly (SAA) is a nearby portion of the Van Allen (radiation) Belts that is
about 200 to 300 km off the southern coast of Brazil.
As the {\em{HST}} orbit precesses and the Earth rotates, the southern part of
the {\em{HST}} orbit encounters the SAA for 7 to 9 orbits and then the next 5 to 6 orbits
(8 to 10 hours) do not intersect the SAA.  During SAA intersections, {\em{HST}} observing
activities must be halted for approximately 20 to 25 minutes \citep{HSTCycle18Primer}.

For observers near the surface of the Earth,
cosmic rays provide an inescapable source of background radiation.
The intensity of cosmic radiation is dependent on altitude, latitude, longitude, azimuth angle
and the phase of the solar cycle. Cosmic ray flux rates are minimum when the Sun's magnetic field is strongest
(during solar maximum) 
and vice versa.

Not all ``cosmic rays'' are of extraterrestrial origin.
Local  ``cosmic rays'' can be generated from particle decay of radioactive materials
near or even {\em{inside}} astrophysical-grade cameras.  
Radiation from concrete telescope piers has been reported\footnote{
``At the ESO 1.52-m telescope on La Silla, we noticed an increased 
`cosmic' ray rate when observing at certain positions on the sky. It turned 
out to be a radioactive `hotspot' in the concrete of one of the telescope 
piers. When the CCD dewar was near this pier, the rate went up.
Concrete is well known to have such hotspots, or concentrations of U and other 
naturally occurring radioactive elements.'' 
Reference: Schwarz, H. E. 2001, CCD-world,\newline
Wed Nov 7 11:08:24 CLST 2001,\newline
http://www.ctio.noao.edu/pipermail/ccd-world/2001/000476.html
}
and recycled-steel rebar, used in the construction of reinforced concrete, can be
radioactive. Radioactive optical elements
used in cameras such as dewar windows made of BK7 glass \citep{Groom2002}
or high-efficiency antireflection coatings made of naturally-radioactive
thorium fluoride can cause a much higher (than expected) incidence of cosmic rays
near the CCD detector. Sometimes even the CCD detectors are themselves
radioactive; some of the early CCD cameras at the European Southern
Observatory (ESO) used thinned backside-illuminated RCA CCDs which were slightly
radioactive
\citep{DodoricoDeiries1987}.

\section{COSMIC-RAY DEGRADATION OF CCD IMAGES}

Cosmic-ray degradation of CCD images is caused by energetic particles passing through
the CCD substrate.  A cosmic ray interacts with the CCD substrate and generates electrons
which are treated by the CCD identically to photoelectrons that are generated from the 
photoelectric effect.

Cosmic rays bombard astrophysical CCD cameras from all angles.  Some cosmic rays
may hit the CCD with an angle of incidence near zero and are seen as small sharp image defects
covering only a few pixels, while others may graze the CCD substrate with 
angles of incidence near 90$\arcdeg$ and are seen as long streaks covering a large number of
pixels as the cosmic ray travels nearly horizontally through the CCD substrate.
While metal shields (typically of tantalum or aluminum) are frequently
placed on the back sides of astrophysical 
cameras to reduce the overall particle flux, the radiation mitigation provided is typically not
much better than
$\sim$2$\pi$ steradians without using unusual shielding configurations.

Since cosmic rays do not go through the optical path of a CCD camera, the
appearance of cosmic ray defects in a CCD image is not blurred by the
Point Spread Function (PSF) of the camera.  As a result,
CCD defects typically have higher spatial frequencies
than is supported by the the optical design of the CCD camera.  Many single-image cosmic-ray
rejection algorithms take advantage of this to separate cosmic-ray defects from 
point-source (e.g., stellar) observations.  Undersampled cameras make the separation
between point sources and vertical CCD defects problematical;
when the undersampling on the focal plane is severe, 
stellar images cover only a few pixels and
vertical cosmic rays appear to look like stars.

Cosmic-ray image defects are not always sharp.
The location where the cosmic ray interacts with the CCD strongly affects the appearance
of the defect  in the CCD image.
Cosmic rays striking horizontally near the surface of a CCD produce sharp-edged streaks 
while those that plough deeply through the CCD substrate produce fuzzy streaks.

\section{COSMIC-RAY REMOVAL IN SINGLE CCD OBSERVATIONS}

An expeditious way  to  reduce the cosmic rays seen in astronomical CCD observations is to
take multiple exposures of the same field and combine the images  by
rejecting  very  high  counts  in  each pixel stack.  The high counts produced
by a cosmic ray will typically be statistically significantly much higher than the 
counts produced by the sky or objects seen in that pixel.
This technique has been well implemented many times
[see, e.g., the
{\sc{IRAF}} \citep{Tody1986spie,Tody1993adass} 
{\sc{STSDAS}} task {\small\tt{crrej}} in the {\small\tt{hst\_calib}} package].
Sometimes, however, it is not possible to obtain multiple exposures due.
Cosmic-ray identification and removal from {\em{single}} CCD observations
is considerably more difficult than with a stack of nondithered CCD
observations.

There are many image-analysis algorithms and applications
available to the astrophysicist for the detection and removal of 
cosmic-ray defects in CCD astrophysical imaging observations.
\cite{FaragePimbblet2005}
have tested the following four commonly used applications for 
cosmic ray reject in single CCD images:
(1) the IRAF script {\small\tt{jcrreg2.cl}}
\citep{Rhoads2000},
(2) the IRAF script {\small\tt{lacos\_im.cl}} \citep{vanDokkum2001}
which is based the {\sc{l.a.cosmic}} algorithm,
(3) the C language program {\small\tt{dcr.c}} 
\citep{Pych2004},
and
(4) Mark Dickinson's IRAF script {\small\tt{xzap.cl}}.
The algorithms used by these four applications are summarized
by \cite{FaragePimbblet2005}; their analysis of the
{\sc{l.a.cosmic}} algorithm follows.
\begin{quote}
``From general observations of the results obtained throughout
the study, the van Dokkum algorithm produces a very
well-cleaned image, though the algorithm tends sometimes
to miss detecting the relatively larger and less elongated
cosmic ray events.''
\end{quote}

\section{L.A.COSMIC ALGORITHM}

The {\sc{l.a.cosmic}} algorithm for cosmic-ray rejection
is based on a variation of 
Laplacian edge detection; it
identifies cosmic rays of arbitrary shapes and sizes
by the sharpness of their edges and can reliably
discriminate between poorly undersampled point sources and cosmic rays.
The {\sc{l.a.cosmic}} algorithm is
described in detail by \citet{vanDokkum2001}.
The process is iterative and typically requires 4 iterations for the
optimal removal of cosmic rays from {\em{HST}} 
Wide-Field Planetary Camera 2 (WFPC2) observations.
Van Dokkum's IRAF script for cosmic-ray rejection in images,
{\small\tt{lacos\_im.cl}}\footnote{The IRAF script
{\tt{lacos\_im.cl}} is
currently available at\newline
http://www.astro.yale.edu/dokkum/lacosmic/lacos\_im.cl}, 
is robust and requires very few user-defined parameters.
The following non-default parameter values work well with low-gain
{\em{HST}} WFPC2 observations:\\[-1.4em]
\begin{itemize}
\itemsep -0.2em
\item
{\small\tt{gain=7}} (gain [electrons ADU$^{-1}]$\,)
\item
{\small\tt{readn=5}} (read noise [electrons]\,)
\item
{\small\tt{sigfrac=0.3}} (fractional detection limit for neighboring pixels)
\item
{\small\tt{objlim=4}} (contrast limit between cosmic rays and an underlying object)
\end{itemize}

Searching the {\em{HST}} Data Archive revealed that the 2400-s CR-SPLIT
WFPC2 observation of the galaxy cluster MS~1137+67 used by van Dokkum
to test {\small\tt{lacos\_im.cl}} was the WF3 section of the WFPC2 dataset
U3060302M.C0H\footnote{IRAF notation: U3060302M.C0H[3]}\,.

Processing that 800$\times$800 pixel
WFPC2 image takes 44 seconds with the IRAF script {\small\tt{lacos\_im.cl}} and the above
non-default parameters running on 
my desktop Apple Mac Pro computer with two 2.8-GHz quad-core Intel Xeons processors.
The run time scales linearly with the number of pixels: processing the upper-right
750$\times$750 pixel subimage\footnote{IRAF notation: U3060302M.C0H[3][51:800,51:800]} 
of that image on the same machine takes 39 seconds [$= 44 * (750^2 / 800^2$)]\,.

\section{CRBLASTER APPLICATION}

Although {\small\tt{lacos\_im.cl}} does an excellent job in removing
cosmic ray defects in WFPC2 images, it has one major drawback -- it is slow.
Why? Because it runs as an IRAF script.
An implementation of the {\sc{l.a.cosmic}} algorithm written in the C language 
should be significantly faster than {\small\tt{lacos\_im.cl}}\,.

Careful reading of \cite{vanDokkum2001} reveals that 
the {\sc{l.a.cosmic}} algorithm is an embarrassingly parallel algorithm.
As such, 
it is ideally suited to being implemented as a parallel-processing 
image-analysis application.

I have written a parallel-processing image-analysis program, 
called 
{\sc{crblaster}},
which does cosmic ray rejection 
of CCD images
using the \mbox{\sc{l.a.cosmic}} algorithm.
\mbox{\sc{crblaster}} is written in C using the high-performance computing
industry standard Message Passing Interface (MPI) library
(e.g., \citeauthor{Snir_etal_1998}\,\citeyear{Snir_etal_1998};
\citeauthor*{GroppLuskSkjellum1998}\,\citeyear{GroppLuskSkjellum1998};
\citeauthor{Pacheco1997}\,\citeyear{Pacheco1997}).
All {\sc{crblaster}} source code and documentation 
with support software and test images are freely available at the official application
website\footnote{{\sc{crblaster}} application website:\newline
http://www.noao.edu/staff/mighell/crblaster}
at the National Optical Astronomy Observatory.

{\sc{crblaster}} uses a two-dimensional (2-D) image partitioning algorithm
which segments an input image into $N$ rectangular subimages of nearly equal area.
{\sc{crblaster}} initially used
a one-dimensional (1-D) image partitioning algorithm
that segmented the input image into $N$ subimages 
that were horizontal slices of the input image of nearly equal area.
The original 1-D partitioning algorithm can be simulated as a 1$\times$N segmentation
with the current 2-D partitioning algorithm.

The {\sc{crblaster}} code has been designed to be used 
by research scientists
as a parallel-processing computational framework
that enables the easy development of other parallel-processing image-analysis programs
based on embarrassingly-parallel algorithms.

\subsection{Single Processor Mode}

Processing the 800$\times$800 pixel test image on my Apple Mac Pro with 
{\sc{crblaster}} running on a single processor
takes $5.95$ 
seconds, which is a speedup factor of 7.40 
over the execution time for the 
{\small\tt{lacos\_im.cl}} 
IRAF script.
This is major progress towards making a faster implementation of the {\sc{l.a.cosmic}} algorithm.

As expected, {\sc{crblaster}} run times with a single processor scale linearly with the number of pixels:
processing the 750$\times$750 pixel test image on the same machine with 
{\sc{crblaster}} running on a single processor
takes 5.23 
seconds
[$= 5.95 * (750^2 / 800^2$)]\,.

Any comparison between two implementations of the {\sc{l.a.cosmic}} algorithm should be done on the parts
of a WFPC2 camera observation that are astrophysically meaningful. 
The left and bottom edges of all WFPC2 PC1, WF2, WF3, and W4 camera observations show
scattered light from the WFPC2's four-faceted pyramid mirror. 
The 750$\times$750 pixel test image is thus better for a comparison than the 800$\times$800 pixel test image 
because that image has pixels on the 
left edge (columns: $x\lesssim35$ px) and the 
bottom edge (rows: $y\lesssim48$ px) 
that show scattered light from the WFPC2 pyramid mirror and {\em{not}} the field of interest.

{\sc{crblaster}} is a high-fidelity implementation in C of the {\sc{l.a.cosmic}} algorithm.
When the output produced by {\sc{crblaster}} and
{\small\tt{lacos\_im.cl}}  using
750$\times$750 pixel test image
 are compared, the inner 786$\times$786 pixels are {\em{identical}};
only 5.11\% of all pixels within 7 pixels of an edge of the image were different.
But why are there any differences at all? Edge effects.
The {\sc{l.a.cosmic}} algorithm, as described in van Dokkum's article or implemented in the 
IRAF script {\small\tt{lacos\_im.cl}},
does not have its behavior explicitly defined along the outer three pixels on each edge of the input image.  
Why the difference in the outer {\em{seven}} pixels?
The largest digital filter used by the {\sc{l.a.cosmic}} algorithm is a 7$\times$7 median filter
whose coding is implementation dependent because its behavior was not explicitly defined within the outer
three pixels of each edge of the input image.

\subsection{Multiple Processor Mode}

An outline of {\sc{crblaster}} in its mutiple-processor mode follows.
The program begins with the initialization of the MPI infrastructure on all nodes (processes).
The director\footnote{I 
use a director/actor paradigm instead of a
master/slave paradigm in describing the choreography of interprocess communication
in a parallel-processing application.  No master willingly acts as a slave 
(except possibly during Roman festival of Saturnalia) --- yet many
directors are actors in their own movies 
(e.g., Charlie Chaplain, Woody Allen, Clint Eastwood, etc.)\,.}  
process then reads the input cosmic-ray-damaged FITS\footnote{
FITS [Flexible Image Transport System: \citet{Wells+1981}\,]
images are read and written using version 3.09
of the {\sc{cfitsio}} library
which is included with the {\sc{crblaster}} software package.
{\sc{cfitsio}}
is currently available at
{{http://heasarc.nasa.gov/docs/software/fitsio/fitsio.html}}~.
}  image from disk and then 
splits it into subimages which are sent to the actor processes.
Each actor process (and sometimes including the director process) then 
does cosmic ray rejection using the {\sc{l.a.cosmic}}
algorithm on its own subimage and when done sends the resulting cosmic-ray
cleaned subimage to the director process.
The director process collects all of the cosmic-ray cleaned
subimages from the actor processes and combines them together 
to form the 
cosmic-ray-cleaned output image which is then
written to disk as a FITS image.
After the program finalizes the MPI infrastructure, {\sc{crblaster}} 
frees up all allocated memory and exits.

After the director process initially
reads the cosmic-ray-damaged input FITS image from disk,
it partitions the input image 
into $N$ rectangular subimages (with overlapping ``edge'' regions)
where $N$ is the number of
processes requested by the user
(i.e., the {\small\tt{np}} parameter value of the {\small\tt{mpirun}} command).
These cosmic-ray-damaged input subimages contain about $1/N^{\rm{th}}$ of the input image
plus an overlap region that is {\small\tt{BORDER}} pixels beyond all joint partition edges. 
For the {\sc{l.a.cosmic}} algorithm, the optimal value of {\small\tt{BORDER}} 
has been determined to be 6 pixels; using less than 6 pixels leaves many cleaning artifacts, 
and using more than 6 pixels does not improve the
quality of the final output image while requiring additional computational overhead.

The director/actor processes then send/receive the cosmic-ray-damaged input
subimages by using {\em{two}} matching pairs of blocking send/receive operations
[{\small\tt{MPI\_Send()}} / {\small\tt{MPI\_Recv()}} calls] for each actor in the order
of the actor's process number [the rank value returned by a {\small\tt{MPI\_Comm\_rank()}} call].

The first pair of  blocking send/receive operations transmits
the contents of the image {\em{structure}} ({\small\tt{struct imageS\_s}})
of the input subimage
as an array of 
{\small\tt{sizeof(struct imageS\_s)}} bytes 
({\small\tt{MPI\_Datatype MPI\_CHAR}})
from the director to the actor. 
A lot of important information about the image {\em{structure}} (but {\underline{not}} the actual image {\em{data}})
of the input subimage is transferred in one operation.
This is a programming hack that greatly simplifies the {\sc{crblaster}} code.
However, this hack does come at a cost of reduced portability: 
{\sc{crblaster}} should only be executed in a {\em{homogeneous}} computing environment.
{\sc{crblaster}} is thus intended for use 
on computer clusters composed of identical CPUs (central processor units)
{\underline{or}} on multi-core machines/servers (e.g., Apple Mac Pros).

The second pair of  blocking send/receive operations transmits
the image {\em{data}} of the input subimages
as an array of {\small\tt{doubles}} ({\small\tt{MPI\_Datatype MPI\_DOUBLE}})
from the director to the actor\footnote{This
works well when the subimage data is being sent to another process. 
If, however, the subimage data is being sent from the director to itself (in its role as an actor), 
then experience has shown that
some implementations of the MPI library may hang or crash
due to assumptions that were made by the creators of that MPI implementation as to
maximum size of message that any sane user would ever wish to self transmit to/from a given process. 
Since the subimage arrays used by {\sc{crblaster}} can easily be many megabytes in size, it is prudent to
replace the {\tt{MPI\_Send()}} / {\tt{MPI\_Recv()}} pair of calls with a simple and fast
{\tt{memcpy()}} memory copy call whenever the director sends subimage data to itself.
}.

Each actor then
does cosmic-ray rejection using the {\sc{l.a.cosmic}}
algorithm on their own cosmic-ray damaged subimage.
The resulting cosmic-ray-cleaned output subimage is the same
size as the input subimage.  
The execution (wall) time required for the actor to complete the 
cosmic-ray rejection task is determined from the difference
between a pair of {\small\tt{MPI\_Wtime()}} calls and 
is recorded in the image {\em{structure}} of the output subimage.

After an actor finishes its task,
it uses a blocking send operation to transmit
the contents of the image {\em{structure}} 
of the output subimage
from the actor to the director
as an array of bytes. 
The director starts with the first actor.
The director uses a {\small\tt{MPI\_Iprobe()}} call to determine if an actor
is ready to transmit its output subimage to the director. 
If an actor is 
(1) {\em{not ready}} (because it is still working),
or
(2) has {\em{already sent}}
the contents of the image {\em{structure}} {\underline{and}} the 
image {\em{data}} of its output subimage 
to the director, then
the director skips that actor and proceeds to the next one.
If an actor {\em{is ready}} to transmit its results to the director,
the director then uses a blocking receive operation to get the contents of the 
image {\em{structure}} of the output subimage.
Once the director receives that transmission (message),
the actor uses a blocking send operation to transmit the image {\em{data}}
of the output subimage from the actor to the director$^9$;
the director uses a blocking receive 
operation 
to get the image {\em{data}} of the output subimage from the actor$^9$ and 
then proceeds to the next actor.
If the last actor has been contacted but there are still actors who have not yet
sent their output subimages to the director, the director goes back to the first actor.
This loop continues
until the director has received all of the output 
subimages from the actors.

The director then 
combines together the {\em{non-overlapping regions}} of the output subimages 
to form the cosmic-ray-cleaned output image.
Finally, the output image is written to disk as a FITS image.

\subsection{Tilera 64-core TILE64 Processor}

I ported {\sc{crblaster}} to the Tilera 64-core TILE64 processor in 8 hours spread over
a few days
using a Tilera TILExpress-20G PCIe card\footnote{A product brief is currently available at\newline
http://www.tilera.com/pdf/Product\_Brief\_TILExpress-20G.php}
installed inside a Dell T5400 workstation
running the CentOS 5.4 implementation of the Linux operating system.

The Tilera 700-MHz TILE64 processor\footnote{The vendor part number is
TLB-26400-2X10-2G-7. A product brief is currently available at 
http://www.tilera.com/pdf/Product\_Brief\_TILE64.php}
on the TILExpress-2G card
features 64 identical processor cores (tiles) interconnected in an 8$\times$8 mesh architecture;
it is programmable in ANSI C and C++ and runs the SMP (Symmetric Multi-Processors) Linux operating system.
Each tile can independently run a full operating system, or a group of multiple
tiles can together run a multiprocessing operating system like 
SMP  Linux.
The TILE64 processor is energy efficient; it consumes 15 to 22 W at 700 MHz
with all cores running full application.
The TILE64 processor has no hardware assist for floating point operations;
all floating point operations are done in software.

\subsubsection{L.A.COSMIC Work Function}

{\sc{crblaster}} processing the 750$\times$750 pixel 
test image on the TILExpress-20G
card using a single processor (tile) took 80.460$\,\pm\,$0.064 seconds (wall time) in 100 trials.
The slowdown factor of 
15.4 between the TILE64 processor
and a Xeon processor on my Apple Mac Pro can be broken down
to a slowdown factor of 4 (due to the difference in clock speeds: 
700 MHz vs.\ 2.8 GHz) multiplied by a 
slowdown factor of 3.8 (due mostly to software emulation of floating point operations).

The initialization and finalization stages of {\sc{crblaster}} (including the reading/writing of the 
input/output FITS images) are the {\em{sequential}} portion of the program
which can not be easily parallelized.
These activities require interaction
with the operating system and spinning physical disk, and consequently
the precise
timing of these activities inevitably varies from one run to another.
The timing variability due to the reading/writing of large images
can sometimes be greatly minimized by reading/writing the image data on/off a 
ramdisk instead of a physical hard disk drive --- reading/writing 
from/to memory can be much faster than
to spinning magnetic disks.

The 100 trials described above took 0.626$\,\pm\,$0.053 seconds
in the intialization and finalization stages; the {\em{parallelizable}}
portion of {\sc{crblaster}} took 79.834$\,\pm\,$0.033 seconds.

{\sc{crblaster}} processing the 750$\times$750 pixel 
test image on the TILExpress-20G
card using 49 tiles took 2.79$\,\pm\,$0.12 seconds (wall time) in 100 trials;
the intialization and finalization stages took 0.78$\,\pm\,$0.12 seconds (wall time)
and the parallelizable 
portion of the application took 2.0118$\,\pm\,$0.0014 seconds (wall time).
The uncertainty of the timing of the sequential portion of the application is
more than 85 times larger than the uncertainty of the parallel portion where
the work of cosmic-ray rejection takes place.

The speedup (factor) of a parallel-processing
computation done with $N$ processes can
be defined as follows,
\begin{displaymath}
{\cal{S}}_N \equiv \frac{t_1}{t_N},
\end{displaymath}
where
$t_1$ is the execution time of the {\em{sequential}} algorithm
and 
$t_N$ is the execution time of the {\em{parallel}} algorithm with $N$ processors.
Ideal (linear) speedup is achieved when $S_N=N$.  
An application with $S_N \approx N$ is considered to have very good scalability.
Computational efficiency is a performance metric 
that can be defined as follows,
\begin{displaymath}
\epsilon
\equiv
\frac{{\cal{S}}_N}{N}.
\end{displaymath}
It typically has a value between zero and one.
Whenever the computational efficiency exceeds one, the parallel algorithm,
by definition,
exhibits superlinear speedups.

The computational efficiencies of {\sc{crblaster}} with the {\sc{l.a.cosmic}} algorithm, 
{\sc{poisson}} (a Poisson noise generator; see \S\,6.3.2), and
{\sc{waiter}} (a microsecond delay function; see \S\,6.3.3)
work functions 
using a 64-core 700-MHz TILE64 processor with 1 to 57 tiles 
is given in Table~1.
In order that the results not be swamped by the noise associated with 
the sequential portion of the application,
the computational efficiencies presented
are for the parallelizable portion of the application (a.k.a.\ inner wall time).
The computational efficiency for these work functions is, by definition, 100\%
for a single processor ($N=1$). 
\begin{center}
\begin{deluxetable}{rrrrr}
\tablecaption{Computational Efficiencies with 3 Work Functions\label{tbl:1}}
\tablewidth{0pt}
\tablehead{
&
\multicolumn{2}{r}{L.A.COSMIC} &
{POISSON} &
{WAITER} 
\\
$N$ &
1-D &
2-D &
1-D &
1-D \\
&
(\%) &
(\%) &
(\%) &
(\%)}
\startdata
1 &  100.00 &  100.00 &  100.00 &  100.00 \\ 
2 &   97.84 &   97.87 &   99.83 &   99.79 \\ 
3 &   95.00 &   94.95 &   99.76 &   99.68 \\ 
4 &   93.59 &   95.70 &   99.47 &   99.36 \\ 
5 &   92.29 &   92.28 &   99.53 &   99.45 \\ 
6 &   90.18 &   92.69 &   99.37 &   99.31 \\ 
7 &   89.41 &   89.43 &   98.65 &   98.56 \\ 
8 &   87.56 &   90.99 &   98.96 &   98.92 \\ 
9 &   86.07 &   89.56 &   98.49 &   98.30 \\ 
10 &   84.95 &   89.05 &   98.91 &   98.85 \\ 
11 &   83.47 &   83.47 &   97.82 &   97.70 \\ 
12 &   82.34 &   87.29 &   98.18 &   98.05 \\ 
13 &   81.01 &   81.02 &   98.31 &   98.11 \\ 
14 &   79.91 &   85.83 &   97.92 &   97.87 \\ 
15 &   79.05 &   85.54 &   98.38 &   98.22 \\ 
16 &   77.90 &   84.42 &   98.34 &   98.10 \\ 
17 &   76.75 &   76.75 &   96.54 &   96.29 \\ 
18 &   76.25 &   83.15 &   97.59 &   97.43 \\ 
19 &   75.10 &   75.08 &   96.97 &   96.81 \\ 
20 &   74.07 &   81.59 &   96.78 &   96.69 \\ 
21 &   72.93 &   87.32 &   97.12 &   97.02 \\ 
22 &   71.92 &   84.74 &   95.24 &   95.11 \\ 
23 &   71.21 &   71.23 &   96.51 &   96.47 \\ 
24 &   70.84 &   87.31 &   95.50 &   95.35 \\ 
25 &   74.58 &   86.76 &   97.05 &   97.00 \\ 
26 &   73.78 &   82.96 &   96.96 &   96.86 \\ 
27 &   72.63 &   84.72 &   96.56 &   96.44 \\ 
28 &   72.24 &   86.13 &   96.35 &   96.28 \\ 
29 &   71.43 &   71.43 &   96.43 &   96.33 \\ 
30 &   70.85 &   85.76 &   96.16 &   96.08 \\ 
31 &   69.05 &   68.98 &   93.45 &   93.47 \\ 
32 &   68.24 &   84.63 &   94.25 &   94.21 \\ 
33 &   68.17 &   82.16 &   95.19 &   95.09 \\ 
34 &   67.99 &   78.40 &   92.33 &   92.17 \\ 
35 &   66.20 &   84.25 &   93.71 &   93.86 \\ 
36 &   66.38 &   83.86 &   95.61 &   95.65 \\ 
37 &   64.95 &   64.93 &   92.93 &   93.06 \\ 
38 &   65.02 &   76.22 &   94.85 &   94.85 \\ 
39 &   63.75 &   79.61 &   92.41 &   92.42 \\ 
40 &   63.82 &   82.40 &   94.44 &   94.60 \\ 
41 &   62.40 &   62.42 &   92.18 &   92.20 \\ 
42 &   62.63 &   82.97 &   94.75 &   94.76 \\ 
43 &   61.31 &   61.26 &   92.38 &   92.49 \\ 
44 &   61.76 &   80.38 &   90.27 &   90.32 \\ 
45 &   60.54 &   81.16 &   93.19 &   93.22 \\ 
46 &   59.25 &   72.21 &   90.72 &   91.15 \\ 
47 &   59.82 &   59.80 &   94.39 &   94.42 \\ 
48 &   58.78 &   80.71 &   92.35 &   92.41 \\ 
49 &   57.66 &   81.01 &   90.40 &   90.44 \\ 
50 &   58.31 &   79.93 &   93.03 &   93.23 \\ 
51 &   57.17 &   75.24 &   91.88 &   92.32 \\ 
52 &   56.19 &   78.16 &   90.43 &   90.34 \\ 
53 &   55.27 &   55.28 &   88.28 &   88.57 \\ 
54 &   55.86 &   78.86 &   92.65 &   92.75 \\ 
55 &   54.84 &   78.27 &   90.60 &   91.04 \\ 
56 &   53.94 &   78.25 &   89.29 &   89.36 \\ 
57 &   53.15 &   72.72 &   87.65 &   87.22 
\enddata
\end{deluxetable}
\end{center}

\vspace*{-2em}
Most of the data in Table~1 is presented graphically in Fig.~1.
\begin{figure*}[th!]
\figurenum{1}
\epsscale{1.0}
\plotone{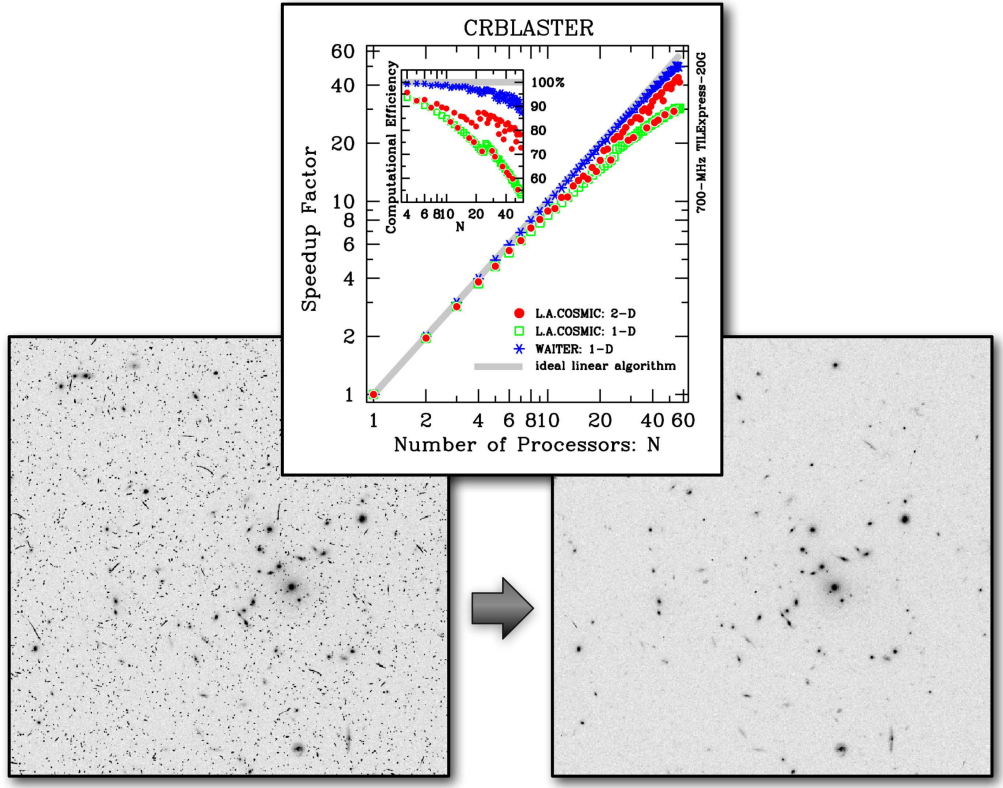}
\caption{The measured performance of CRBLASTER with 1 to 57 processors using a  
64-core 700-MHz TILE64 processor (see Table~1).}
\end{figure*}
The left image 
in the figure is part of a {\em{HST}} WFPC2 observation 
of the galaxy cluster MS 1137+67 (i.e., the 750$\times$750 pixel test image).
The application of the {\sc{l.a.cosmic}} algorithm to the input image produces the 
image on the right.  Both images were plotted with the same lookup table
to simplify comparisons.
Note that most 
--- but not all --- of the cosmic rays seen in the input image have been removed.
The large graph in Fig.~1 shows the speedup
factor as a function of the number of processors on a log-log plot
and the inset graph 
shows the computational efficiency
as a function of the number of processors on a log-linear plot.
The performance of the {\sc{waiter}} work function is shown with asterisks ($*$) 
and the performance of {\sc{l.a.cosmic}} work function is shown with 
open squares (1-D image partitioning)
and filled circles (2-D image partitioning).
The performance of an ideal linear algorithm is shown,
for comparison purposes, as a grey straight line
in both graphs.  
The performance of the {\sc{poisson}} and {\sc{waiter}}
work functions (columns 4 and 5 of Table~1, respectively) 
are so similar that only the performance data
for the {\sc{waiter}} work function is shown in Fig.~1.

For the {\sc{l.a.cosmic}} algorithm,
the closer the subimages are to being square, the more efficient the computations
will be (for any given number of processors and a square input image).  
Whenever the number of processors
used is a prime number the 2-D image partitioning is least efficient ($1/N$);
note that the computational efficiency values for the {\sc{l.a.cosmic}} algorithm
are nearly identical in columns 2 and 3 of Table~1 for the following prime number
values of $N$: 
2, 3, 5, 7, 11, 13, 17, 19, 23, 29, 31, 37, 41, 43, 47, 53.
This effect can also be seen in Fig.~3 where the filled circles (2-D data)
are plotted on top of the open squares (1-D data)
whenever the number of processors is a prime number.

Edge effects directly impact computational efficiency.
Analyzing the 750$\times$750 pixel test image with a 1$\times$36 segmentation
produces many subimages with a total of 24,750 pixels ($=33\times750$); 
21 rows of 750 pixels width
is the actual subimage data to be analyzed
and 2$\times${\small\tt{BORDER}} rows of 750 pixels width are the edge data
above and below the actual subimage data.  The ratio between edge pixels and the
actual image pixels is 57.1\% ($=12/21$).
Analyzing the same image with a 6$\times$6 segmentation
produces many subimages with a total of 18,769 pixels ($=137^2$); 
the ratio between edge pixels and the
actual image pixels is a significantly better 20.1\% [$=(137^2-125^2)/125^2$].
The time to analyze a typical 1$\times$36 subimage was
25.2\% slower 
than the time to analyze a typical 6$\times$6 subimage
(3.18 s versus 2.54 s wall time).
This slowdown reflects directly on the computational efficiency:
Table~1 shows that the computational efficiency for 
the {\sc{l.a.cosmic}} work function for 36 processors
is  66.38\% and 83.86\%, respectively,
with 1-D and 2-D input image segmentation ($1.2633= 83.86/66.38$).

Van Dokkum (\citeyear{vanDokkum2001}) claimed and 
I showed in \S5 and \S6.1 that the
run time of the {\sc{l.a.cosmic}} algorithm
scales linearly with the number of pixels.
However, that statement is not always true.  A better statement would be
as follows: the run time of the {\sc{l.a.cosmic}} algorithm
scales linearly with a {\em{large}} number of pixels.
In this context, $750^2$ is a large number of pixels.

A detailed analysis of the {\sc{l.a.cosmic}} algorithm
shows that algorithm is a {\em{nonlinear}} process in that 
the amount of computation required for any given pixel
{\em{is not constant}}:
pixels in or near a cosmic-ray get more attention
(take longer to analyze)
than pixels in a part of the input image that 
has not been corrupted by cosmic-rays.

Cosmic-ray damage of a CCD observation is a random process;
the longer the exposure, the greater the probability increases that any given pixel
will be corrupted.  So when a cosmic-ray damaged CCD observation is broken down into
a large number of subimages, one of the subimages will have the most
cosmic-ray damage and one will have the least.

When the 750$\times$750 pixel test image is split into 36 subimages
with a 6$\times$6 segmentation,
there are 16 subimages of size $137^2$ pixels 
which have typical minimum and maximum execution (wall) times of
1.90 and 2.54 seconds, respectively, with 
the {\sc{l.a.cosmic}} work function; the median was 2.52 seconds.
The slowest actor took 33.7\%  more time to execute
than the fastest actor. {\sc{crblaster}} must wait until the slowest actor has finished 
its cosmic-ray rejection task before it can produce the clean output image.
This unavoidable
waiting for results lowers of the computational efficiency of the application.

\subsubsection{POISSON Work Function}

The computational efficiency of the
underlying computational framework of {\sc{crblaster}}
can be determined by replacing
the {\sc{l.a.cosmic}} work function
with a light-weight (low memory usage) embarrassingly-parallel
algorithm that is a {\em{linear}} process
that has a equal computation load for each pixel in the input image.
The replacement work function
should not require any overlap regions for the input/output subimages 
({{\small\tt{BORDER=0}}).

I have developed a nearly-linear alternate work function called {\sc{poisson}}
which is based on a simple yet wonderfully inefficient Poisson noise generator,
{\small\tt{Poisson\_fnI1}} (see Fig.~ 2), 
that is my C language implementation
of Algorithm Q 
in {\em{Seminumerical Algorithms}} \citep{Knuth1969}.
While Algorithm Q is simple, its 
complexity grows linearly with the mean of the Poisson distribution ($\mu$);
it runs quickly with small mean values but slowly with large mean values ($\mu\ga1000$).
The {\sc{poisson}} work function uses the value of an input subimage pixel
as the mean of the desired Poisson distribution 
from which a random Poisson deviate is drawn; the value of the deviate
is then
stored in the associated
pixel of the output subimage\footnote{Code snippet: 
{\tt{out[y][x] = Poisson\_fnI1( in[y][x] )}}}.
The {\sc{poisson}} work function does not use any overlap regions for the input/output
subimages ({{\small\tt{BORDER=0}}).
\begin{figure}
\figurenum{2}
\setlength\fboxrule{2pt}
\hspace*{2.5truemm}
\fbox{\plotone{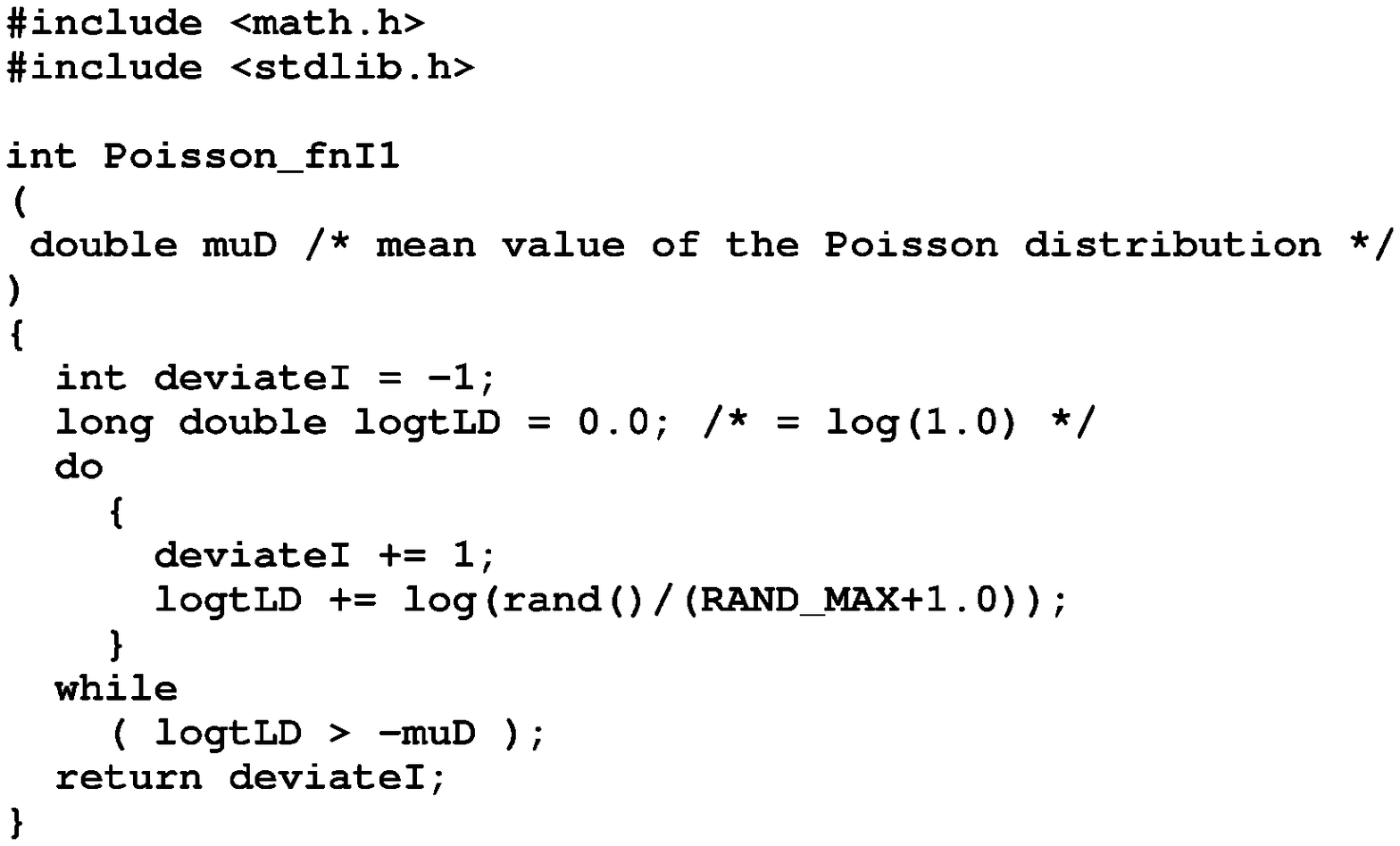}}
\caption{An inefficient Poisson noise generator.}
\end{figure}

A 750$\times$750 pixel test image for {\sc{poisson}} was created,
{\footnotesize\tt{in\_750x750\_24.3.fits}}, with every
pixel value set to 24.3.
That value was chosen to give the {\sc{poisson}}
work function an inner wall time for a single tile that is 
approximately equal to the inner wall time of the {\sc{l.a.cosmic}} work function
for a single tile with the 750$\times$750 test image.

{{\sc{crblaster}} using the {\sc{poisson}} work function
processing the {\footnotesize\tt{in\_750x750\_24.3.fits}} test image 
on the TILExpress-20G
card using a single tile took 
79.738$\,\pm\,$0.016
seconds (inner wall time) in 100 trials.
Table~1 shows the computational efficiency of the
{\sc{poisson}} work function for 1 to 57 tiles.
The computational efficiencies given in Table~1 
for {\sc{poisson}} are for 1-D image segmentation; the values
for 2-D image segmentation are nearly identical --- that is not surprising 
considering that the work function does not use overlap regions in the input/output subimages.

The underlying computational framework of {\sc{crblaster}}
is efficient.  {\sc{crblaster}} using the {\sc{poisson}}
work function has a computational efficiency of 95.61\% with 36 tiles
and 90.40\% with 49 tiles (see Table 1).  

\subsubsection{WAITER Work Function}

The {\sc{poisson}} work function is a {\em{nearly}} linear process because 
since it uses
a Poisson noise generator, the execution time for each pixel is {\em{almost}}
but not {\em{exactly}} the same.  

I have developed a {\em{linear}} alternate work function called {\sc{waiter}}
that does nothing but waits for $X$ microseconds 
where $X$ is the value of the input subimage pixel.  
{\sc{waiter}} does not require the use of any overlap regions for the input/output
subimages ({{\small\tt{BORDER=0}}).

A 750$\times$750 pixel test image for {\sc{poisson}} was created,
{\footnotesize\tt{in\_750x750\_137.fits}}, with every
pixel value set to 137.
That value was chosen to give the {\sc{waiter}}
work function an inner wall time for a single tile that is
approximately equal to the inner wall time of the {\sc{l.a.cosmic}} work function
for a single tile with the 750$\times$750 test image.

{{\sc{crblaster}} using the {\sc{waiter}} work function
processing the {\footnotesize\tt{in\_750x750\_137.fits}} test image 
on the TILExpress-20G
card using a single tile took 
79.7202$\,\pm\,$0.0021
seconds (inner wall time) in 100 trials.
Table~1 and Fig.~1 shows the computational efficiency of the
{\sc{waiter}} work function for 1 to 57 tiles.
The computational efficiencies given in Table~1 
for {\sc{poisson}} are for 1-D image segmentation; the values
for 2-D image segmentation are nearly identical because
the work function does not use overlap regions in the input/output subimages.

{\sc{crblaster}} using the {\sc{waiter}}
work function has a computational efficiency of 95.65\% with 36 tiles
and 90.44\% with 49 tiles (see Table 1). 

Sometimes the {\em{nearly}} linear 
{\sc{poisson}} work function is {\em{more}} efficient than 
linear {\sc{waiter}} 
work function (see Table 1).  How can that be? 
{\sc{crblaster}} is optimized to have the director
receive output subimages whenever they are available\footnote{{\sc{crblaster}}
deliberately does not use the {\tt{MPI\_Barrier}}
command.
Indiscriminate use of the
{\tt{MPI\_Barrier}}
command
can be 
an excellent way to {\em{lower}} the computational efficiency of a parallel-processing
image-analysis application like {\sc{crblaster}}.}. 
That way when the slowest
tile has finished its job, all other results have probably already been received by the director.
In the case
of the {\sc{waiter}} work function most of the big tiles finish at approximately the same
time --- that can cause a small delay in processing due to the backlog of results that the director
needs to process at same time.

\subsection{Apple 8-core Mac Pro}

{\sc{crblaster}} with the {\sc{l.a.cosmic}} work function
processing the 800$\times$800 pixel test image with a 2$\times$4 segmentation
on my Apple Mac Pro using all 8 cores
takes $0.875$ 
seconds (wall time), which translates to a computational efficiency 85.0\% for 8 processors;
this is a speedup factor of 50.3 times faster than the IRAF script running on one processor.
The same experiment with the
750$\times$750 pixel test image 
takes $0.749$ 
seconds (inner wall time)
which is a computational efficiency 86.9\% for 8 processors;
this is 4.1\% lower than the 2-D value of 91.0\% of the TILE64 processor with 8 tiles.
The small difference between the computational efficiencies for the Apple Mac Pro with
two quad-core Xeons and 64-core TILE64 processor
is not at all surprising considering their very different processor architectures and
clock speeds.

\section{DISCUSSION}

Hardware configurations
(processor type and speed, available memory, interconnect speed, disk drive speed, etc.)
will cause different clusters and
machines to see different computational efficiencies with {\sc{crblaster}}.
For example, a cluster using a Fast Ethernet (100 Mbits s$^{-1}$) interconnect
will see lower computational efficiencies 
when doing cosmic-ray rejection on {\sl{HST}} WFPC2 observations with
{\sc{crblaster}} 
than if it had a Gigabit Ethernet (1 Gbits s$^{-1}$)
interconnect.

Can the implementation of the {\sc{l.a.cosmic}} algorithm in {\sc{crblaster}}
be improved?  Certainly.  
The current C code of {\sc{crblaster}} is almost literally a line-by-line translation
of van Dokkum's IRAF script {\small\tt{lacos\_im.cl}}; it is  
a high-fidelity implementation of that script 
--- warts and all.

The underlying computational framework of {\sc{crblaster}} could be improved
by making the two-dimensional image partitioning algorithm more efficient.
For example, the 2-D image partitioning algorithm for 37 processors currently
uses a 1$\times$37 image segmentation.  
One way that image partitioning
could be done more efficiently would be to do a 6$\times$6 segmentation 
and then split one of the subimages in half; the computational efficiency
would then be much closer to that of the 2-D  6$\times$6 value 
than the 1-D 1$\times$37 value (see Table~1).

The underlying computational framework of {\sc{crblaster}} could be improved
by further minimizing the time expended doing message passing.  
The current implementation is pretty efficient but it might be possible to eliminate
a few tens of milliseconds during the message passing stage of the application ---
but the price would likely be high: a considerable amount of coding effort would probably
be required at the cost of making the code probably much more complicated.

If one does not care to know how long each actor took to execute its 
cosmic-ray rejection task, then a few milliseconds could be saved
by simply not transmitting the output subimage {\em{structure}} contents
since the contents of the input
and output subimages are identical except for the actor execution time information.

The {\sc{crblaster}} code can be used as a software framework for easy development of 
parallel-processing image-anlaysis programs using embarrassing parallel algorithms.  Two 
alternate work functions ({\sc{poisson}} and {\sc{waiter}}) have been 
provided as examples on how to implement embarrassing parallel algorithms
within the {\sc{crblaster}} computational framework.
The biggest required 
modification to the {\sc{crblaster}} code
is the replacement of the core image-processing 
work function with an alternative work function that is a {{sequential}} implementation
of an embarrassing-parallel algorithm. 
If the new algorithm needs an 
overlap region of the subimages, then the numerical value of {\small\tt{BORDER}}
will need to be modified to the appropriate value for the new algorithm.
And of course, the command line options will need to be modified to provide the new algorithm 
information about any custom user-supplied parameters.
Beyond these simple modifications, nothing else
within the main software framework needs to be touched.

The information sent by the actors to the director
does not have to be the same type of information
that the director sent to the actors.
For example, instead of transmitting cosmic-ray-cleaned output subimages, the actors could
send back arrays of structures containing the
stellar photometry of stars in the input subimage --- if one replaced
the {\sc{l.a.cosmic}} work function with a sequential stellar photometry engine.
The programmer would, of course, need to write new communication functions using
MPI to properly send/receive the new information correctly 
to/from the director/actors. The current {\sc{crblaster}} code has been written
in a pedagogical fashion such that the creation of new communication functions
should be a relatively simple effort for programmers who are knowledgeable in the
C language but are not expert MPI programmers.

In order to take full advantage of the {\sc{crblaster}} computational framework, one should have the actors
spend much more time working than communicating with the director.  This can be done by having the actors
do very time-consuming tasks that take many minutes or hours with a single processor --- tasks like galaxy/stellar photometry.  It might be possible to port the {\sc{crblaster}} computational framework
to existing complex tasks like the {\sc{MultiDrizzle}}\footnote{Main {\sc{MultiDrizzle}} website: \newline
http://stsdas.stsci.edu/multidrizzle/}
software package.  That, however,
would likely be a challenging port in that {\sc{MultiDrizzle}} is written in Python and is intended to 
run within a PyRAF environment; it might very well be possible to do 
but a great deal of thought would likely be required in order to develop
the proper interfaces for cleanly calling Python {\sc{MultiDrizzle}} 
code from the C functions within {\sc{crblaster}}.  Porting C applications would clearly be a much
simpler proposition.

If {\sc{crblaster}} were to be considered
for doing onboard cosmic-ray rejection on a future NASA astrophysical imaging
mission, then further optimization of the application would be highly
recommended in order to make the most efficient use of onboard computational facilities.

\acknowledgements

I am grateful to Dagim Seyoum for loaning me a TILExpress-20G card along with 
the supporting software.
I thank Alan George and the staff of the 
High-performance Computing \& Simulation Research
Lab (University of Florida) for providing access
to the Sigma cluster during the early stages of this research.
I would also like to
thank the following people for providing stimulating conversations and advice
during the port of {\sc{crblaster}} to the 64-core TILE64 processor: Vijay Aggarwal,
Marti Bancroft,
Norm Burke,
Steve Crago,
Rob Kost,
Michael Malone,
John Samson,
and 
Jinwoo Suh.
This work has been supported by a grant from the National Aeronautics and
Space Administration (NASA), Interagency Order No.~NNG06EC81I 
which was
awarded by the Applied Information Systems Research
(AISR) Program of NASA's Science Mission Directorate.


\end{document}